\documentclass[pdflatex,sn-mathphys-num]{sn-jnl}

\usepackage{graphicx}%
\usepackage{multirow}%
\usepackage{amsmath,amssymb,amsfonts}%
\usepackage{amsthm}%
\usepackage{mathrsfs}%
\usepackage[title]{appendix}%
\usepackage{xcolor}%
\usepackage{textcomp}%
\usepackage{manyfoot}%
\usepackage{booktabs}%
\usepackage{algorithm}%
\usepackage{algorithmicx}%
\usepackage{algpseudocode}%
\usepackage{listings}%
\usepackage{mdframed}
\usepackage{multicol}

\theoremstyle{thmstyleone}%
\theoremstyle{thmstyletwo}%

\theoremstyle{thmstylethree}%

\begin{document}

\title{Practical Application and Limitations of AI Certification Catalogues in the Light of the AI Act}


\author*[1]{\fnm{Gregor} \sur{Autischer}}\email{gregor.autischer@student.tugraz.at}

\author[2]{\fnm{Kerstin} \sur{Waxnegger}}\email{kwaxnegger@know-center.at}

\author*[1,2]{\fnm{Dominik} \sur{Kowald}}\email{dkowald@know-center.at}

\affil[1]{\orgname{Graz University of Technology}, \orgaddress{\city{Graz}, \country{Austria}}}

\affil[2]{\orgname{Know Center Research GmbH}, \orgaddress{\city{Graz}, \country{Austria}}}


\abstract{In this work-in-progress, we investigate the certification of AI systems, focusing on the practical application and limitations of existing certification catalogues in the light of the AI Act by attempting to certify a publicly available AI system. We aim to evaluate how well current approaches work to effectively certify an AI system, and how publicly accessible AI systems, that might not be actively maintained or initially intended for certification, can be selected and used for a sample certification process. Our methodology involves leveraging the Fraunhofer AI Assessment Catalogue as a comprehensive tool to systematically assess an AI model's compliance with certification standards. We find that while the catalogue effectively structures the evaluation process, it can also be cumbersome and time-consuming to use. We observe the limitations of an AI system that has no active development team anymore and highlighted the importance of complete system documentation. Finally, we identify some limitations of the certification catalogues used and proposed ideas on how to streamline the certification process.}




\maketitle

\section{Introduction}\label{sec1}

Artificial intelligence (AI) has evolved, over several decades, to complex machine learning (ML) algorithms and neural networks that are common-place today. However, in recent years, AI systems were increasingly integrated into our daily lives, moving from specialized research labs to mainstream applications \citep{costa_applications_2023}. Today AI is utilized in many  domains such as healthcare  \citep{rajpurkar2022ai}, human resources \citep{van2021machine}, finance \citep{cao2022ai} as well as in prominent technologies such as recommender systems for e.g., social networks \citep{kowald2013social,lacic2014recommending} or online marketplaces \citep{lacic2014towards,lacic2015utilizing}. AI now plays a significant role in shaping our interactions with technology and informing decision-making processes across various sectors \citep{kasinidou_artificial_2024}. The rapid proliferation of AI applications has raised concerns about safety, privacy, fairness, and further ethical implications \citep{baum_fear_2023,kowald2024establishing}.  
In response to these challenges, AI governance has become an increasingly prominent focus for legislators and policymakers worldwide \citep{smuha_race_2021}. The European Union's AI Act, for instance, represents a landmark piece of legislation that aims to establish a comprehensive regulatory framework for AI systems \citep{mueck_etsi_2022}. Similar initiatives are underway in other countries and regions. This reflects a growing global consensus on the need for AI governance \citep{nad_trustworthiness_2023}. Organizations and policymakers are increasingly emphasizing the need for AI system certification to achieve regulatory compliance and foster confidence in AI systems. Since the EU AI Act entered into force on August 1, 2024, this is now more important than ever.
The certification of AI systems is a complex challenge, distinctly different from traditional software certification. While conventional software certification primarily focuses on functionality and security, AI certification must address a broader spectrum of concerns, including prediction accuracy, fairness, transparency, and other ethical considerations \citep{winter_trusted_2021}.

With this work-in-progress, we aim to bridge the gap between theoretical certification frameworks and their practical application. We attempt to certify an existing open-source AI system using current certification frameworks and document the challenges encountered along the way. We primarily use the Fraunhofer AI Assessment Catalogue published in  \citet{poretschkin_ai_2023}. However, we draw comparisons to other catalogues. With this approach, we aim to:
\begin{itemize}
\item Identify which parts of the catalogue are most useful and if simplifications or refinements could be beneficial.
\item Provide a more practical understanding of the potential AI certification process.
\item Discover the limitations where sample certifications encounter challenges.
\end{itemize} 
Therefore, we provide a comprehensive walk-through of an AI certification in practice. Our work starts with essential background information on selected AI regulations and existing certification catalogues. We then detail the specific AI system chosen for certification, and the core of this work documents our AI certification attempt. We offer an analysis of what aspects of certification were achievable and which proved problematic. We conclude this paper by describing these findings and by offering recommendations for future developments in AI certification schemes.

\section{Current State of AI Regulation}
\label{sec2}

Rapid advancement and widespread adoption of AI in various industries require the development of comprehensive regulatory frameworks, and increasing AI deployment in more critical areas appears to require stricter regulation \citep{pimentel_why_2024}. Some bodies like the US Food and Drug Administration have already approved certain AI applications, particularly in medicine \citep{benjamens_state_2020}. In the European Union, some medical applications also got approval, one example is the ChestLink software, that automatically reports chest x-rays that are classified as normal. Systems such as this set an important precedent for other medical and potentially nonmedical systems in the future \citep{saenz_autonomous_2023}. Recently, lawmakers have recognized the urgent need for comprehensive regulation. New regulatory frameworks intend to encompass not only individual industries, but also to create universal rules that ensure consistency and fairness. Worldwide, several organizations are advancing policy initiatives \citep{nad_trustworthiness_2023}. Now, we describe some key efforts in this area.

\subsection{EU Artificial Intelligence Act}

The European Union's AI Act is currently the most significant and far-reaching regulatory initiative in the field of AI, and it took effect on 1st of August 2024. The impact of the AI Act will extend far beyond the EU's boarders, potentially setting global standards for AI management.
The AI Act uses a risk-based approach to achieve the following goals \citep{european-union_regulation_2024}:
\begin{itemize}
    \item Ensure AI safety and compliance: guarantee that AI systems in the EU are safe and adhere to laws protecting fundamental rights and EU values, safeguarding users' privacy and preventing discrimination.
    \item Certainty for AI investment: establish a clear legal framework to foster innovation and investment in AI, providing businesses and investors with regulatory clarity.
    \item Enhance governance: strengthen enforcement mechanisms to effectively apply existing laws on fundamental rights and AI safety requirements.
    \item Unify AI market: foster a single, cohesive market for trustworthy AI applications across the EU, preventing fragmentation through harmonized regulations.
\end{itemize}

\subsubsection{Scope}
The AI Act covers various actors and scenarios within the AI ecosystem \citep{european-union_regulation_2024}. The regulation applies to providers, deployers, importers, and distributors of AI systems, as well as product manufacturers incorporating AI systems into their products, regardless of their location, if the AI system or its output is used within the European Union. The AI Act explicitly excludes, among others, AI systems developed purely for scientific research and development. It also excludes AI systems published under open-source licences. However, the regulation applies if an organisation brings an open-source system to market or uses it as a prohibited AI system, a high-risk AI system or an AI system with special transparency obligations. The AI Act imposes obligations across the entire AI value chain, ensuring a comprehensive approach to AI governance and safety within the European market.

\subsubsection{Definition of AI}
The AI Act adopts a broad and technology-neutral definition of AI systems, focusing on their functional characteristics rather than specific technologies or methods. According to the AI Act, an AI system is defined as a machine-based system designed to operate with varying levels of autonomy, which potentially exhibits adaptiveness after deployment, and that is capable of generating outputs such as predictions, content, recommendations, or decisions that can influence physical or virtual environments based on input it receives \citep{european-union_regulation_2024}. This definition emphasizes two key elements: 'inference' and 'autonomy', which distinguish AI systems from traditional software with predetermined outputs. The AI Act takes a broad approach to stay relevant as technology rapidly advances. It encompasses both the core AI system and its surrounding code, recognizing the complexity of AI applications. This definition aligns with recent conceptualizations of AI and moves away from earlier, more restrictive definitions tied to specific technologies \citep{fernhout_eu_2024} (e.g., ML). The AI Act creates a framework that can accommodate current and future AI technologies, ensuring its long-term applicability in regulating the AI landscape.

\subsubsection{AI Risk Categories}
The AI Act categorizes AI systems into various risk levels and imposes corresponding requirements, with stricter regulations for higher-risk applications \citep{european-union_regulation_2024}. Key categories include:
\begin{enumerate}
    \item Prohibited AI systems.
    \item High-risk AI systems.
    \item General-purpose AI systems.
    \item AI systems with special transparency obligations.
    \item Limited-risk AI Systems.
\end{enumerate}
Prohibited systems include, but are not limited to, systems that manipulate behaviour, exploit vulnerabilities, or create facial recognition databases from untargeted scraping. The AI Act defines high-risk AI systems as those that pose significant risks to health, safety, or fundamental rights, and that are either used as products (or components of products) covered by specific EU legislation and require third party certification, or that are listed in Annex III of the AI Act \citep{european-union_regulation_2024}. Annex III includes several areas such as biometric identification, emotion recognition systems, management of critical infrastructure, education, employment, and law enforcement. General-purpose AI systems face some regulation, but it is less stringent than for high-risk systems. The AI Act also provides an exception for certain AI systems that, despite falling under Annex III categories, may not be considered high-risk. This is the case if they perform narrow procedural tasks or do not significantly influence human decision-making, provided they do not involve profiling of natural people. AI systems with special transparency obligations, according to Article 50, require clear disclosure when users interact with AI or encounter AI-generated content. Limited-risk AI systems are only subject to voluntary codes of conduct for ethical and responsible use,  according to Article 95.

\subsection{EU Artificial Intelligence Liability Directive}
Complementing the AI Act, the European Commission proposed the AI Liability Directive in September 2022. This directive aims to modernize and enhance the EU's liability framework for AI systems \citep{madiega_artificial_2023}, and also to ensure that individuals who suffer damages from AI systems receive equivalent protection to those harmed by other forms of technology. By standardizing liability regulations across the EU, the directive aims to avoid legal discrepancies and guarantee uniform protection for those impacted by AI-related damages. The AI Act and other EU measures coordinate with the AI Liability Directive, which addresses only non-contractual liability claims, encompassing a wide array of potential AI-related harms \citep{european-union_proposal_2022}. This comprehensive strategy seeks to balance the protection of victims with the encouragement of AI innovation, reducing legal uncertainties and promoting the responsible advancement of AI technologies within the EU.

\subsection{Other Global Initiatives}
While the EU has developed one of the most comprehensive regulatory efforts, other countries and regions have also proposed AI regulations. The following examples represent some relevant international regulatory efforts. For instance, the US has discussed a Blueprint for an AI Bill of Rights, which outlines principles for the design and deployment of AI systems, although they currently only have a patchwork of state laws \citep{white-house_blueprint_2022}. The UK has discussed a sector-led approach to AI regulation, and Japan has developed AI Guidelines emphasizing a multi-layered governance framework \citep{department_for_digital_culture_media__sport_establishing_2022}. These initiatives do not have legal enforceability. They still highlight the global acknowledgment of the necessity for AI regulation and certification efforts. The current state of AI regulation evolves rapidly, with the EU taking a leading role by introducing the AI Act.

\section{Certifying AI}
\label{sec3}
Organizations use certification as an important and established tool to prove that technical systems meet certain standards or regulations. Successful certifications play a vital role in proving a system's compliance with applicable norms. It also plays a crucial role in establishing trust in a system among its users \citep{blosser_consumer_2023}. For traditional software projects, where every block of code can undergo review line by line, companies have long-established certification processes \citep{winter_trusted_2021}. 

Organizations and regulatory bodies are still in the early stages of certifying AI applications. This is partly because policymakers and other actors have just recently begun developing comprehensive legal frameworks for AI. A prime example is the aforementioned EU AI Act. Despite ongoing efforts by international standardization organizations like ISO, IEC, and IEEE to create guidelines and standards, they have yet to fully establish a certification process \citep{nad_trustworthiness_2023}. These efforts aim to address the distinct challenges presented by AI technologies. However, the absence of comprehensive legal frameworks has hindered the development of robust certification processes for AI. Even with recent legislative developments, the rapidly evolving nature of AI technology continues to pose significant challenges for creating and maintaining effective certification standards \citep{delgado-aguilera_jurado_introduction_2024}. This dynamic landscape requires certification processes that are both adaptable and rigorous, capable of evolving alongside the technology they aim to regulate.

Certifying AI systems also presents unique challenges due to their complex and often opaque nature. Often, humans do not directly program logical rules to model decision-making processes. Instead, AI systems use various methods to analyse and interpret data, learning  their own rules for decision-making processes. Humans do have multiple ways to influence and understand how the system makes decisions and how good the outcome is. However, the process lacks complete transparency. This complexity necessitates a different approach to certification, emphasizing the need to establish a comprehensive framework that accounts for the inherent opacity and adaptability of AI technologies \citep{winter_trusted_2021}. 
Moreover, AI systems can evolve and change their behaviour over time by retraining with new data. This continuous learning process adds another layer of complexity to certification, as regulators and developers must continuously monitor and re-evaluate systems to ensure they comply with ethical standards and avoid biases. Addressing these dynamic aspects is essential to develop robust certification practices that can keep pace with the rapid evoluation of AI \citep{benjamin_fresz_contribution_2024}.

\subsection{AI Certification Catalogues}
CEN, CENELEC and ETSI are leading European standardization bodies, they bridge the gap between EU regulations and practical certification frameworks designed to evaluate and certify AI systems. They integrate these guidelines with European legislative priorities, and ensure consistency across the European standardization landscape \citep{hadrien_pouget_standardsetzung_2024}. Several organizations have created catalogues and guidelines to evaluate, test, and certify AI systems. Prominent examples include the Fraunhofer AI Assessment Catalogue by \citet{poretschkin_ai_2023}, the white paper 'Trusted Artificial Intelligence' by \citet{winter_trusted_2021}, and the white paper 'Auditing Machine Learning Algorithms' by the supreme audit institutions of various countries \citep{sai-fi-de-nl-no-uk_auditing_2023}. These frameworks provide distinct methodologies and list criteria to certify AI applications, addressing aspects such as fairness, autonomy and control, transparency, reliability, safety and security, and data protection. With this work, we focus primarily on the Fraunhofer Certification Catalogue and how it applies to a concrete AI application.

\subsubsection{Fraunhofer AI Assessment Catalogue}
The Fraunhofer AI Assessment Catalogue emphasizes the necessity of implementing stringent quality standards to ensure AI systems are reliable, safe, aligned with societal values and compliant with the law, particularly in sensitive application contexts \citep{poretschkin_ai_2023}. The catalogue identifies several key challenges in assessing and ensuring AI quality. These include the complex value chain involved in AI development and the difficulty in explaining the inner workings of AI models. The authors of the catalogue argue that these challenges necessitate a systematic approach to quality implementation in AI development and highlight the importance of unbiased expert assessment in establishing trust in AI applications.
A significant focus of the catalogue lies in operationalizing quality requirements for AI. While there are established guidelines for reliable AI, the catalogue points out that the specifics of their practical application remain largely unclear. The paper proposes a risk-based AI assessment approach and introduces an AI assessment catalogue. This catalogue provides a structured approach for certifying AI applications across different dimensions of trustworthy AI: fairness, autonomy and control, transparency, reliability, safety and security, and data protection. The proposed framework offers a procedure for developing safeguarding arguments for AI applications. It tries to support developers and operators of AI systems in meeting regulatory requirements. It introduces a step-by-step process that this work uses when trying to certify an AI System. 

\subsubsection{Trusted Artificial Intelligence}
This white paper published by T\"UV Austria and Johannes Kepler University Linz wants to outline a structured approach to certifying ML applications \citep{winter_trusted_2021}. It describes key ML principles and discusses relevant aspects and challenges in the context of certification. It emphasizes that while ML systems are complex, they are not black boxes, but rather white boxes whose operations can be analysed in detail. The paper introduces a certification approach for ML applications, focusing initially on supervised learning tasks with low-risk potential. It focuses mainly on technical aspects and cybersecurity measures. The paper presents a summary of the certification framework. However, since the actual catalogue remains inaccessible to the public, auditors cannot use it directly to certify an AI model. Instead, it serves as a reference point, highlighting key areas to consider during the certification process.

\subsubsection{Auditing Machine Learning Algorithms}
A collaboration of European public auditing institutions released this white paper. It highlights the increasing use of AI and ML in public services, emphasizing the need for new certification methodologies \citep{sai-fi-de-nl-no-uk_auditing_2023}. It identifies several risks, including an over-focus on numerical metrics at the expense of compliance and fairness, miscommunication between product owners and developers, over-reliance on external expertise, and uncertainty regarding personal data use. To address these challenges, the paper proposes a certification framework covering the entire AI application lifecycle. The audit areas focus on data understanding, model development, performance, and ethical considerations such as explainability and fairness. To aid in this process, the paper introduces a helper tool, in the form of a spreadsheet. Auditors can use it to prepare and conduct AI audits efficiently. The authors stress that specialized knowledge and skills are required for ML certifications. They emphasize that the proposed audit catalogue and helper tool should be continuously refined and updated. Ultimately, this paper aims to provide guidance and good practices to enable auditors to navigate different parts of the certification process.

\section{Our AI Application: Facial Emotion Recognition}
\label{sec4}
To undertake a certification process, the first requirement is a system that either requires certification or is eligible for it. The system should incorporate an AI component, ideally leveraging ML techniques. Furthermore, the AI component must be integrated into a larger, comprehensive system, as certification typically applies to entire systems rather than isolated components. Specifically, the Fraunhofer Certification Catalogue mandates a well-defined assessment object, which requires the AI component to be part of a larger, integrated system \citep{poretschkin_ai_2023}.

Addressing these challenges requires the auditor to identify and select a suitable system that meets these criteria. The system should include a ML-based AI component, and should demonstrate sufficient complexity and integration to justify certification. The objective is to find a system where the AI component plays a critical role in the overall functionality, thereby making the certification process relevant and meaningful. 
In essence, our aim is to identify a system that incorporates a ML-based AI module within a broader architecture.

\subsection{Our Decision to Use the EmoPy Framework and RIOT Project for AI Certification}
After careful consideration, we decided to use the EmoPy Framework \citep{angelica_perez_thoughtworksartsemopy_2021} and its implementation within the RIOT Project \citep{thoughtworks_riot_2018} for this sample certification. Several key factors influenced this choice, to ensure that both the framework and the project are suitable for the certification process. 
Firstly, the EU AI Act's relevance significantly influenced our decision. Emotion recognition, the primary focus of the EmoPy Framework, aligns with the potential coverage of the EU AI Act. While open-source models like EmoPy are generally exempt, this system may fall under the AI Act's scope if it was brought to market as a high-risk AI system, a prohibited AI system or an AI system with special transparency obligations.

Another critical factor was the open-source nature and transparency of EmoPy. EmoPy is fully open-source, enabling an in-depth look into the technical details and making the system fully transparent, which should make the certification process possible. The codebase of EmoPy is relatively small and manageable, making it easier to understand and verify. Yet, it is sufficiently large to make a certification worthwhile. Additionally, the framework includes thorough documentation, with multiple articles and resources that describe the model selection process. This level of documentation is critical for reproducibility attempts of the AI models \citep{semmelrock2023reproducibility,semmelrock2024reproducibility}, and with this also for the certification process, as it provides clear insights into the design and functionality of the model, facilitating a thorough evaluation. 
While solid documentation is essential for any certification, it is especially critical for our sample certification, since no company or active development team is managing the project anymore. Although no active development team provides ongoing support, contacting authors and lead developers of the EmoPy framework articles and code, proved beneficial. They kindly addressed questions about the framework and the RIOT setup. Beyond this input, we needed to extract all necessary information from the provided documentation. This reliance on static resources poses a limitation compared to standard certification processes, where ongoing interactions with active developers should be possible. 

From a technical perspective, we considered the suitability of the EmoPy framework for sample certification a key factor. Its technical characteristics and comprehensive documentation make it well-suited for the certification process. Additionally, the integration of EmoPy within the RIOT project provides a complete system context, which is essential for certification. Traditional certification catalogues often struggle to validate standalone ML models. However, the RIOT project offers a comprehensive framework where the AI component is embedded within a broader system. This integration is critical, as certification typically applies to an entire system rather than individual components. 
All these factors influenced our choice to use the EmoPy framework and the RIOT project for our sample certification. Key considerations included its potential relevance to the EU AI Act, its open-source nature ensuring transparency, and its suitability for a sample certification. Additionally, the comprehensive system context that the RIOT project provides reinforced this decision, enabling an effective and thorough certification process. 
It is also important to note that both EmoPy and RIOT were not originally intended for certification by their creators. However, they serve well for this academic exercise, as the focus is on the certification procedure rather than the system's quality or real-world applicability.

\subsection{Brief Overview of the EmoPy Framework}
The EmoPy framework provides multiple neural network architectures for facial expression recognition, including ConvolutionalNN, TransferLearningNN, and ConvolutionalLstmNN. These architectures vary in complexity, with the ConvolutionalNN beingthe simpleste, and TransferLearningNN (using Google's Inception-v3) being the most complex. The authors experimented with different architectures and found that the ConvolutionalNN provides the best overall performance \citep{perez_emopy_2018}. The EmoPy documentation suggests using two publicly available datasets for training and evaluation: the Microsoft FER2013 dataset and the Extended Cohn-Kanade dataset. The key goals of the EmoPy project are to provide free, open-source, and easy-to-use facial expression recognition capabilities, and to advance research in this field by making the models and datasets publicly available. The Framework was also used for the emotion recognition in the RIOT Project.

\subsection{Brief Overview of the RIOT Project}
\begin{figure}[t!]
\centering
\includegraphics[width=0.7\textwidth]{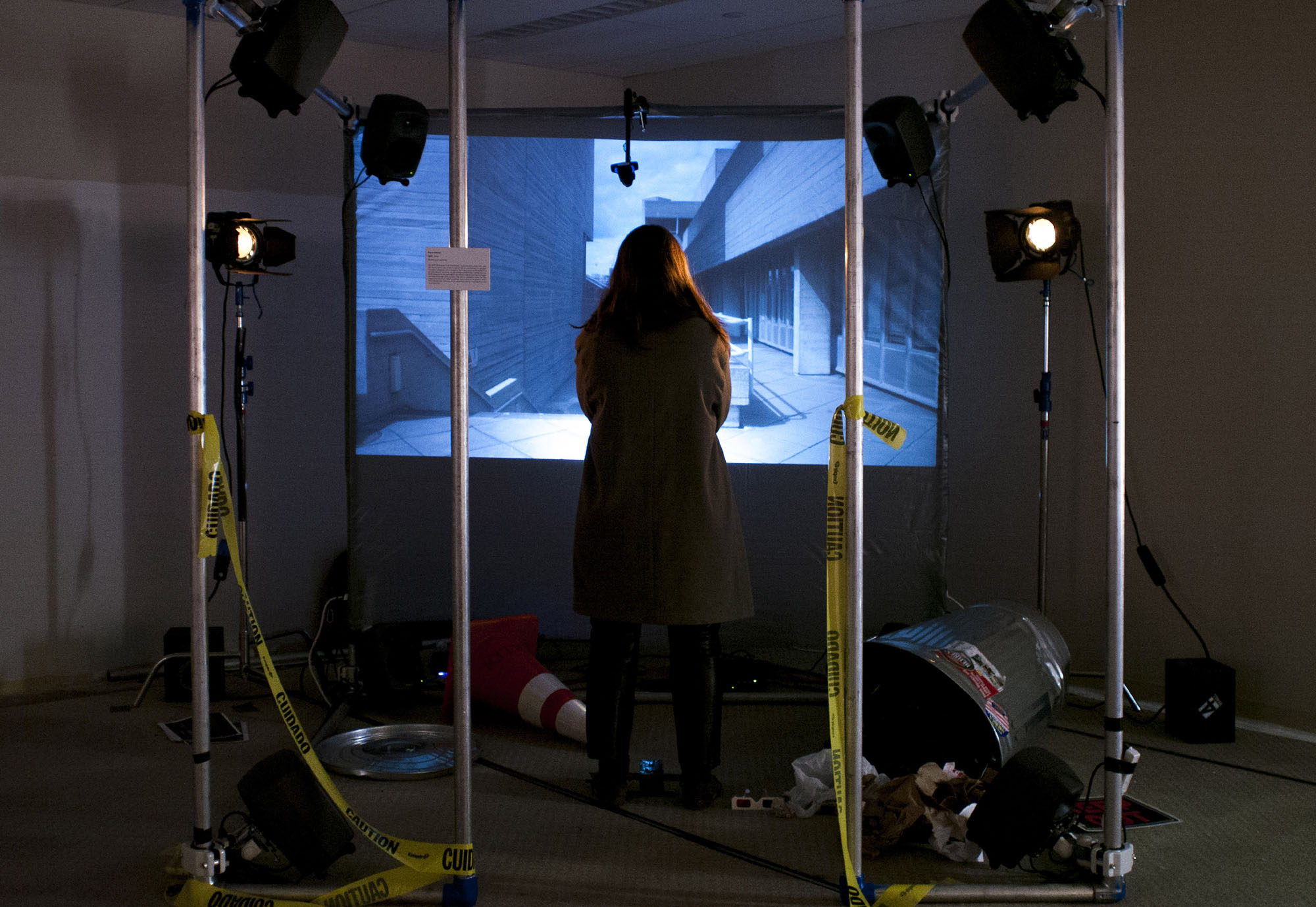}
\caption[RIOT installation]{\textbf{RIOT installation}. This image shows the RIOT Art installation in New York City in 2018. One participant is standing in front of the screen taking part in the experience \citep{perez_emopy_2018}.\vspace{-2mm}} 
\label{fig:riot_installation_example}
\end{figure}

In a nutshell, RIOT is a live-action film that dynamically responds to emotions, utilizing facial emotion recognition technology to guide viewers through an ongoing dangerous riot. The experience allows the audience's emotions to drive the narrative of the film in real-time \citep{thoughtworks_riot_2018}. The project began during artist Karen Palmer's 2017-2018 residency at Thoughtworks, where she developed a new iteration of the emotion analysis engine and the RIOT user experience. The RIOT installation has since been showcased at various events and festivals.

The RIOT experience integrates the EmoPy framework and its pretrained emotion recognition model into its system to respond to participants' emotional states during the live-action film sequence. The characters and narrative adapt to the viewer's detected emotions, creating an immersive, multisensory experience that, according to its creators, enhances cognitive skills and self-awareness \citep{palmer_riot_2016}. The system setup is shown in Figure~\ref{fig:riot_installation_example}. A participant stands in front of the screen to watch the experience. At different intervals, the mounted webcam captures the person's face and predicts their emotion. Based on the detected emotion, the film progresses differently, creating an interactive experience \citep{perez_emopy_2018}. 

\subsection{The AI Application to be Certified}
For this sample certification, the RIOT installation provides the complete system context for the certification process. The EmoPy framework, which provides ML-Models to detect emotions, serves as the core framework within this installation. Our certification process focuses on validating the AI system within the RIOT context. The surrounding code, setup, and information that comprise the entire art installation are relevant, as we cannot certify the AI system independently of these components. However, this approach has shortcomings, as not all the required information is available. Our approach allows for an exploration of the certification process while acknowledging its limitations and academic nature. To facilitate the certification process, we prepared a complete overview of all available information about the AI application, which is summarized below. We also compiled a complete technical overview of the system, that we certified. 

\subsubsection{Compilation of System Resources}

    \textbf{RIOT Installation}
    \begin{itemize}
    \item GitHub repository of the RIOT art installation \citep{thoughtworks_thoughtworksartsriot_2019}
    \item Article on the RIOT art installation \citep{thoughtworks_riot_2018}
    \item TED Talk on the RIOT art installation \citep{ted_residency_karen_2018}
    \item Article that describes different art installations (one of them is the RIOT art installation) \citep{thoughtworks_thoughtworks_2018}
    \item Article on Karan Palmer (the artist behind the RIOT Art Installation) \citep{thoughtworks_karen_2017}
    \item Short description of the RIOT art installation by Karan Palmer \citep{palmer_riot_2016}
    \item Video that showcases and describes the RIOT art installation \citep{karen_palmer_riot_2017}
\end{itemize}

    \noindent \textbf{EmoPy Framework}
    \begin{itemize}
    \item GitHub repository of the EmoPy Framework \citep{angelica_perez_thoughtworksartsemopy_2021}
    \item Article describing the EmoPy Framework and technical decisions that were made in more detail \citep{perez_emopy_2018}
    \item Article describing in more detail decisions that were made regarding the architecture of the Emotion recognition model \citep{perez_recognizing_2018}
    \item Python documentation of the EmoPy Framework \citep{emopy_development_team_welcome_2017}
    \item GitHub repository of the FER+ dataset \citep{microsoft_microsoftferplus_2023}
    \item Extended Cohn-Kanade dataset \citep{cohn_resources_2024}
    \end{itemize}

  \noindent \textbf{The AI Application that is Certified} 
    \begin{itemize}
    \item The AI Application is not the entire RIOT art installation, but only the system subset that receives centred images from the webcam and returns the emotion predictions.
    \item Since the pretrained model and the target emotions used during training are not disclosed, we assume that the EmoPy ConvolutionalNN model is used without any alterations. Furthermore, we assume that the target emotions are anger, fear, calm, and surprise. We presume that the model with these features is the model used in the installation.
    \item We treat all items listed above as if they are part of a proper documentation of the system. We also treat items that are articles or videos as part of the documentation resources that we can use for the AI certification process.
    \end{itemize}

\subsubsection{Technical Overview}
The AI Application that we attempted to certify has the following specifications:
The developers trained the ConvolutionalNN from the EmoPy Framework, as this architecture had the best performance during early testing. The layers and parameters of the ConvolutionalNN can bee seen in Table~\ref{tab:model_updated_summary}.

\begin{table}[h]
    \centering
    \begin{tabular}{|l|l|l|r|}
        \hline
        \textbf{Layer Name}     & \textbf{Type}         & \textbf{Output Shape}       & \textbf{Parameters} \\ \hline
        conv2d\_1              & Conv2D                 & (None, 61, 61, 10)          & 170                 \\ \hline
        conv2d\_2              & Conv2D                 & (None, 58, 58, 10)          & 1610                \\ \hline
        max\_pooling2d\_1      & MaxPooling2D           & (None, 58, 29, 5)           & 0                   \\ \hline
        conv2d\_3              & Conv2D                 & (None, 55, 26, 10)          & 810                 \\ \hline
        conv2d\_4              & Conv2D                 & (None, 52, 23, 10)          & 1610                \\ \hline
        max\_pooling2d\_2      & MaxPooling2D           & (None, 52, 11, 5)           & 0                   \\ \hline
        flatten\_1             & Flatten                & (None, 2860)                & 0                   \\ \hline
        dense\_1               & Dense                  & (None, 4)                   & 11444               \\ \hline
    \end{tabular}
    \caption{Summary of model layers and parameters. \citep{perez_emopy_2018} \vspace{-7mm}}
    \label{tab:model_updated_summary}
\end{table}
In total the model has 15640 trainable parameters and has 4 outputs to be able to classify 4 emotions. The model architecture consists of multiple convolutional and max-pooling layers, which gradually reduce the spatial dimensions while extracting hierarchical features. A final dense layer connects the extracted features to the output layer for classification of the emotions: anger, fear, calm or surprise.

To train the model, two datasets were used. The FER2013 dataset contains over 35,000 facial expression images across 7 emotion classes, while the Cohn-Kanade dataset includes 327 facial expression sequences. However, for this application only the subset of four emotions (anger, fear, calm and surprise) were used. The number of images for each of these emotions can be seen in Table~\ref{tab:emotion_comparison}. To increase the size and suitability of the training and validation datasets, the authors applied data augmentation techniques. The training process begins by splitting the dataset into training and validation subsets. During training, the network weights are iteratively adjusted to minimize the loss between predicted and labelled emotions. To mitigate overfitting, the authors monitored the gap between training and validation accuracy. This approach prevents overfitting and ensures the model generalizes well by using unseen validation data during training. They measured performance using training and validation accuracy, to help refine the models \citep{angelica_perez_thoughtworksartsemopy_2021}.

\begin{table}[h]
\centering
\begin{tabular}{|l|c|c|}
\hline
\textbf{Emotion} & \textbf{FER2013 Quantity} & \textbf{Cohn-Kanade Quantity} \\ \hline
Anger            & 4953                        & 45                     \\ \hline
Fear             & 5121                        & 25                     \\ \hline
Calm             & 6198                        & 0                      \\ \hline
Surprise         & 4002                        & 83                     \\ \hline
\end{tabular}
\caption{Emotion distribution comparison between FER2013 and Cohn-Kanade datasets for used emotion subset \citep{perez_emopy_2018}. \vspace{-7mm}}
\label{tab:emotion_comparison}
\end{table}

\begin{figure}[H]
\centering
\includegraphics[width=0.65\textwidth]{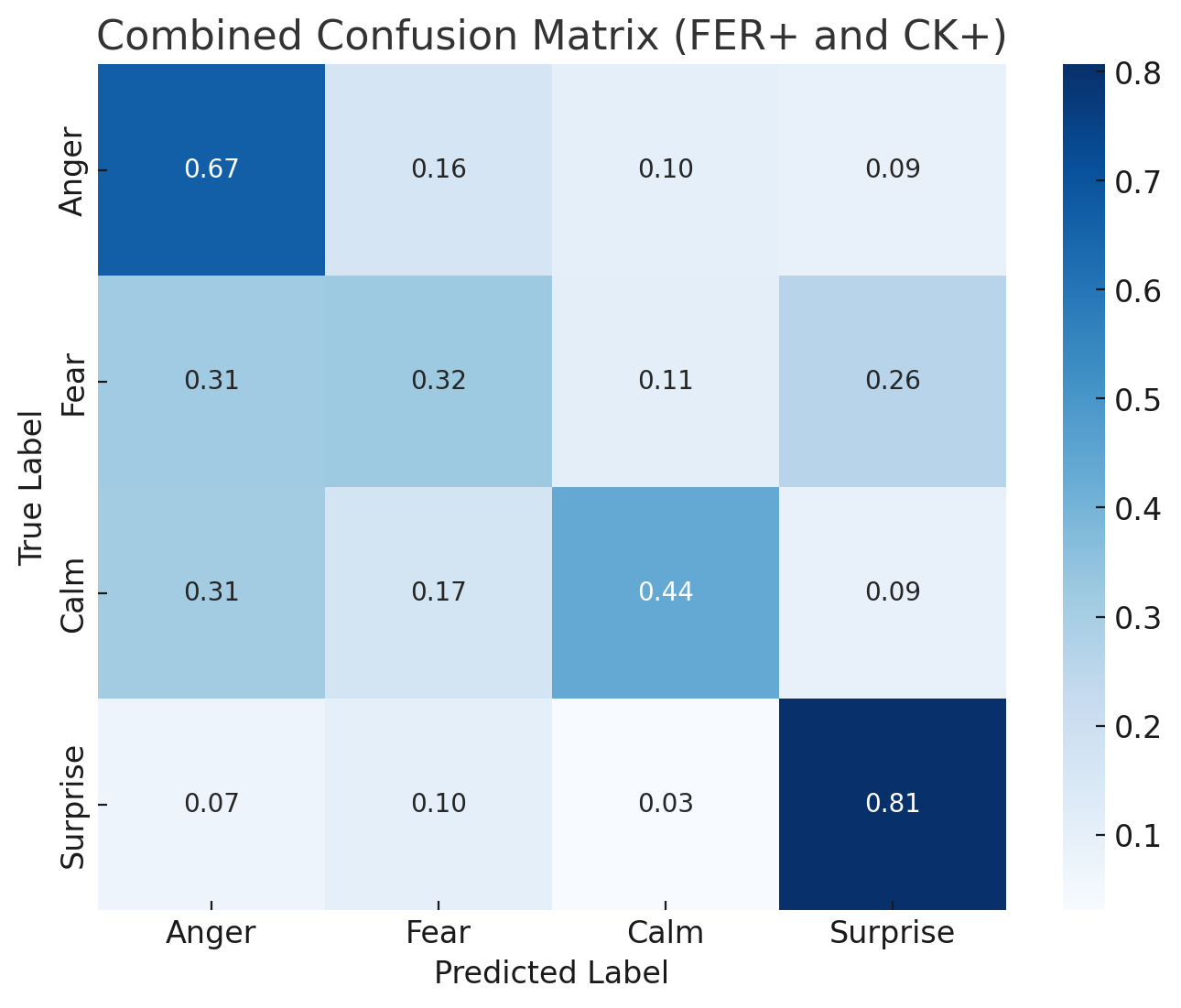}
\caption[Confusion matrix of the trained ConvolutionalNN]{Confusion matrix of the trained ConvolutionalNN \citep{perez_emopy_2018}. \vspace{-5mm}} 
\label{fig:confusion_matrix}
\end{figure}

The developers tested the model and also analysed the confusion matrices to gain insights into the model's performance \citep{perez_emopy_2018}. Some confusion matrices are also published \citep{angelica_perez_thoughtworksartsemopy_2021}. The ones relevant to this application with the used emotion subset for both the FER2013 and the Cohn-Kanade Dataset are also published, and for brevity we combined them into one confusion matrix that represents the AI-systems performance for our certification attempt. We considered that the Cohn-Kanade dataset is approximately 100 times smaller than the FER2013 dataset. The resulting matrix can be seen in Figure~\ref{fig:confusion_matrix}. The data used during training to create this matrix consists only of the discussed datasets (FER2013 and Cohn-Kanade). However, the final model was trained on these datasets with additional data generated through data augmentation. 

\section{AI Certification Approach}
\label{sec5}
We apply the Fraunhofer AI Certification Catalogue to an existing AI system to explore and evaluate the certification process. We chose the facial emotion recognition component of the RIOT art installation, which uses the EmoPy framework. We selected this system because it is open-source, appears well-documented, and is integrated into a larger application context. Additionally, as we have discussed previously, it is potentially covered by the EU AI Act. In this work-in-progress, we use the Fraunhofer Catalogue as the primary certification framework due to its comprehensive nature and full public availability. The catalogue provides a structured approach to AI certification, addressing multiple dimensions of risk. The Fraunhofer Catalogue outlines the certification process, which involves these key steps \citep{poretschkin_ai_2023}:

\begin{enumerate}
\item \textbf{First Step:} get an overview of the System (AI Profile (PF)) and define the AI-System and the boundaries to the surrounding system.
\item \textbf{Second Step:} define the life cycle of the AI application.
\item \textbf{Main Step:} get an overview over all the risk dimensions.
\begin{enumerate}
\item \textbf{Protection requirements analysis:} determine which risk dimensions apply.
\item \textbf{Risk Analysis:} for each applicable dimension:
\begin{enumerate}
\item Risk analysis and objectives.
\item Criteria for achieving objectives.
\item Measures.
\item Overall assessment of a risk area.
\item Summary of each dimension.
\item Cross-dimension assessment.
\end{enumerate}
\end{enumerate}
\item Drawing conclusions and making a certification decision based on the success of the cross-dimensional assessment.
\end{enumerate}

After completing the certification process with the Fraunhofer Catalogue, we analyse and address several key aspects of the process, and we draw conclusions about the challenges we encounter. In this paper, we explore the challenges of selecting an appropriate AI application for certification. We also evaluate how effectively the Fraunhofer certification process worked for this specific case, and highlight its strengths and areas for improvement. Furthermore, we present a comparative analysis, examining how the other two introduced catalogues differ from the Fraunhofer approach and how they could potentially enhance or complement the certification process. Lastly, we evaluate the limitations of this approach. Our analysis addresses the constraints of the chosen AI system and certification catalogue, as well as the applicability of the findings to other AI applications and certification scenarios. With this comprehensive evaluation, we aim to provide valuable insights into the practical implementation of AI certification processes and contribute to the ongoing discourse on AI certification.

\subsection{Before the Certification Process}
Before starting the certification process, we completed several preparatory steps. First, we selected the AI application. We forked the GitHub repository of the EmoPy project and identified the correct dependencies to enable a detailed examination of the codebase. We conducted an extensive research phase, focusing on both the EmoPy framework and the RIOT installation. This research resulted in the creation of the system summary, as previously presented. We created this basis, which combines elements of both the EmoPy framework and the RIOT installation, to facilitate the certification process. This step was necessary to provide a complete system context for certification, as certifying standalone ML models can be challenging with the Fraunhofer certification catalogue and would differ significantly from a real-world certification.
During the preparation phase, we identified certain information gaps in the original documentation. To make the certification process feasible, we closed these gaps with some assumptions and additional details. We made these additions, based on reasonable interpretations of the available information and common practices in AI development, and kept them to a minimum to enable the certification process.

\subsection{AI Profile}
The Fraunhofer Catalogue outlines the first formal step in the certification process: completing the AI Profile. This step was straightforward due to the thorough research we conducted in the preparation phase. The AI Profile offered a structured overview of the system's functionality, intended application context, and key characteristics.

\subsection{Life Cycle of the AI Application}
Following the AI Profile, we conducted the life cycle overview. Although not explicitly stated as a distinct step in the Fraunhofer Catalogue, we found it beneficial to gain a thorough understanding of the AI system's development and operation stages. The AI life cycle encompasses all the stages an AI system undergoes, from planning and development to deployment, operation, ongoing maintenance, and potentially continued model training, ensuring trustworthiness and compliance throughout its use. We adapted the questions for this life cycle overview from a table in the Fraunhofer Catalogue, covering aspects such as data acquisition, model development, and operational considerations.
        
\subsection{Protection Requirement Analysis}
The protection requirement analysis serves as an important first step in the certification process, identifying the risk dimensions that require more in-depth analysis. This analysis involves evaluating the potential impact of the AI system across various dimensions such as fairness, reliability, and data protection.
We examined all dimensions and identified several with medium risk. A summary of the Protection Requirement Analysis can be seen in Table~\ref{tab:protection_requirements_analysis}. For the purposes of this work, we selected two of the required risk dimensions, namely reliability and fairness, for detailed analysis. This selection helps us to focus the sample certification and manage the scope of the certification process. Exploring these two dimensions is also sufficient to understand the certification procedure and draw the appropriate conclusions.

\begin{table}[h]
\centering
\begin{tabular}{|l|l|p{6cm}|}
\hline
\textbf{Dimension}        & \textbf{Risk Category} & \textbf{Reasoning}                                                                 \\ \hline
Fairness                  & Medium                 & Processes personal data (facial images), output linked to personal behaviour but has limited impact on rights. \\ \hline
Autonomy and Control      & Medium                 & Temporarily influences user perception during the interactive film experience.         \\ \hline
Transparency              & Low                    & Lack of transparency does not affect safety or usefulness significantly.               \\ \hline
Reliability               & Medium                 & Misclassification can degrade user experience but does not cause major harm.           \\ \hline
Safety and Security       & Low                    & No risk of physical injury or significant financial damage from malfunction.            \\ \hline
Data Protection           & Medium                 & Uses personal data (live facial images), though not particularly sensitive or secret data. \\ \hline
\end{tabular}
\caption{Protection Requirements Analysis Summary}
\label{tab:protection_requirements_analysis}
\end{table}

\subsection{Risk Analysis}
The risk analysis forms the core of the certification process. We carried out this step by working through a questionnaire from the Fraunhofer Catalogue. The questions addressed different aspects of the selected risk dimensions. For each dimension, we covered topics such as data quality, model design, testing procedures, and operational considerations. A summary of the conclusion of our Risk Analysis for our investigated dimension can be seen in Table~\ref{tab:risk_analysis}. Following the individual dimension analyses, we conducted a cross-dimensional assessment to identify potential trade-offs or interactions between the examined dimensions. This step is crucial to ensure a complete understanding of the AI system's performance and risks. In our sample certification, we were unable to conclude the certification attempt with a positive result, but found significant shortcomings of the AI Application.

\begin{table}[h]
\centering
\begin{tabular}{|l|p{8cm}|}
\hline
\textbf{Dimension} & \textbf{Summary of Risk Analysis} \\ \hline
Reliability        & The system performs well enough within its defined scope, but lacks complete documentation and robustness testing. It is certifiable with improvements in testing and documentation. \\ \hline
Fairness           & Insufficient analysis of potential biases and discriminatory behaviour. No clear metrics or target groups defined for fairness. Not certifiable without significant further improvements. \\ \hline
\end{tabular}
\caption{Certification Summary for Evaluated Dimensions}
\label{tab:risk_analysis}
\end{table}

In the following section, we discuss in detail the challenges we encountered during this certification process. Based on these experiences, we draw conclusions. Additionally, we discuss comparisons with two other certification catalogues to provide a broader perspective on AI certification methodologies.

\section{Results and Main Findings}
\label{sec5}

We performed the certification of the chosen facial emotion recognition system. The core of the certification process with the Fraunhofer Catalogue involves the Protection Requirement Analysis and the Cross-dimensional assessment. Performing the certification allows us to highlight the challenges that arise from using this certification approach in general. It also brings attention to issues that make a work like this more difficult, such as finding a suitable AI system to certify in the first place. This work did not prove that the system complies with today's regulations, nor was that its intent. Our certification process itself did not cover all the necessary dimensions required to certify the system fully. The conclusion of the certification suggests that even the dimensions we closely examined were not sufficient for a successful certification. In a real certification scenario, we would communicate these shortcomings to the development team, so they can address the issues and allow the system to be certified. If full certification were feasible, the process would demonstrate the AI system’s compliance with specific standards outlined in the Fraunhofer Catalogue. We describe all the challenges and potential improvements found in the following paragraphs.

\subsection{Selecting the AI System}
In a conventional certification scenario, the process of selecting an AI system for certification is usually not a consideration, as the system to be certified is predetermined by the organization seeking certification. However, for the purposes of this work, the selection of an appropriate AI system represented a crucial first step that significantly influenced the subsequent certification process and what can potentially be learned from it. The selected system provided a mostly robust foundation for the certification effort. Its existing application context, and good documentation of both the AI model and its surrounding system, enabled meaningful progress through the certification process. During the certification process, we saw the importance of considering both technical factors and solid documentation when choosing an AI system for certification.

\subsubsection{Initial Considerations and Challenges}
The selection of an appropriate AI system requires careful consideration of multiple interconnected factors. While an initial approach might suggest identifying and selecting a standalone AI model or neural network, this proves insufficient when considering the comprehensive requirements of a certification processes. Particularly, the Fraunhofer Catalogue, which we primarily used in this paper, takes an extensive look at how to define the system and its boundaries with other software components. This definition cannot be found with a standalone AI model.

\subsubsection{Context and Embedding Requirements}
The certification process inherently demands a broader contextual framework than what might be immediately apparent. Rather than existing in isolation, the AI system must be embedded within a larger operational context and demonstrate clear use case applications. While it would be theoretically possible to construct artificial use cases for the certification purpose, such an approach could result in a suboptimal certification scenario.
In this paper, therefore, we have chosen an AI application, which has already  established a real-world application context. This characteristic proved invaluable, as it enabled a more natural translation into a certifiable system, providing the necessary surrounding information to support a comprehensive certification approach. The existence of this practical context significantly enhanced the certification process's authenticity and relevance.

\subsubsection{Documentation Requirements}
Documentation emerged as another critical factor in the selection process, on two distinct levels. First, the AI model itself must be thoroughly documented, providing technical specifications and operational parameters. Second, and equally important, the surrounding system infrastructure must be comprehensively documented to make certification feasible. This documentation requirement significantly narrows the field of suitable candidates for certification studies.
A particular challenge encountered in this work relates to the absence of a development team. When selecting an existing model for certification, there is typically no active development team invested in the certification process. This situation creates a significant constraint. Without the ability to request additional documentation or engage in an iterative process with developers, the available documentation must be sufficiently comprehensive from the outset. Any information gaps that arise during the certification process cannot be supplemented or clarified.

\subsection{The Certification Process}
We used the Fraunhofer catalogue as the primary basis for this certification due to its comprehensive and detailed nature. Although we considered other catalogues, such as the TÜV catalogue, they presented significant limitations. The incomplete publication of the TÜV catalogue made it unsuitable for use in the certification process. The Auditing Machine Learning Algorithms catalogue, while fully published and potentially suitable for certification, employs a substantially different approach compared to the Fraunhofer catalogue. Its less step-by-step nature potentially presents additional challenges for those with limited certification experience.

\subsubsection{Main Challenges During the AI Certification Process}
\paragraph{The system's documentation}
We encountered several key challenges during the certification process. One fundamental challenge is that the system's documentation was not originally intended for certification purposes. Additionally, the development process did not require extensive and detailed documentation. This limitation created occasional gaps in documentation that would be essential for a complete certification. In some instances, we made adequate substitutions beforehand to create a more realistic certification scenario. The documentation of a system is key to certification. Our choice of a publicly available system that was not intended for certification has its shortcomings. This choice makes a sample certification, such as the one we attempted here, more difficult and potentially less meaningful.

\paragraph{No active development team}
The absence of an active development team emerged as a critical limitation in the certification process. Without ongoing development support, we could not implement the typical feedback loop, where certification findings would normally lead to documentation improvements and system adjustments. In a standard certification scenario, identified gaps or shortcomings trigger an iterative process of enhancement, with the development team actively working to make the system more certifiable. However, in this paper, we had to evaluate the system purely based on its existing documentation and state. Therefore, we could only have two possible outcomes: either certifiable or not certifiable with the available materials. 

This limitation became more complicated by the fact that the system was originally developed several years ago, and the entire development team had moved on from the project. The lead developer generously provided time to answer questions. However, because the project is old, certain details became less accessible or clear over time. This combination of inactive development and the system being old created a static evaluation scenario, rather than the dynamic, iterative process that typically characterizes successful certification efforts where development teams actively work towards certification compliance.

\subsubsection{Specific Observations on the Fraunhofer Catalog}
The implementation of the Fraunhofer catalogue revealed several notable characteristics and challenges. The catalogue's documentation-centric approach makes it nearly impossible to use for code-only projects, as it focuses exclusively on documentation rather than direct code examination. While code can inform the certification process and documentation creation, the catalogue never directly addresses or describes code.
The catalogue's high specificity and detail provide comprehensive coverage, reducing the likelihood of overlooking critical aspects. However, this thoroughness occasionally results in similar or nearly duplicate questions, increasing the time required for certification completion. The strong documentation focus means less direct attention to mathematical or technical system operations. While the neural network structure remains important for documentation purposes, the Fraunhofer Catalogue only requires it to be examined implicitly rather than explicitly. This approach can be advantageous when dealing with proprietary information, as documentation alone might suffice for a potential certification. A particular strength of the Fraunhofer catalogue lies in its clear differentiation between the AI model, system, and embedding code, which proves crucial in determining certification scope and requirements. This distinction helps ensure appropriate certification coverage.

\paragraph{T\"UV Catalogue}
One notable limitation of the Fraunhofer catalogue is the lack of guidance on how to answer the posed questions. In this regard, a more technology-centric catalogue like the one from T\"UV could provide valuable complementary guidance. The T\"UV catalogue, despite its publication limitations and restricted focus on ML and supervised learning systems, offers useful insights into the technical aspects of ML systems operations.

\paragraph{Auditing Machine Learning Algorithms Catalogue}
The Auditing Machine Learning Algorithms Catalogue presents a markedly different structural approach compared to Fraunhofer's. Its topic-based organization consolidates related questions – for instance, grouping all data-related questions together – contrasting with Fraunhofer's distributed approach where data-related questions appear across various subsections. This structural difference complicates the potential combination of these catalogues. However, the auditing catalogue's reduced duplication could potentially streamline the certification process.

\subsection{Learnings and Recommendations}

The certification process revealed several significant insights regarding both methodological approaches and practical certification challenges. 

\subsubsection{Catalogue-specific Observations}
The Fraunhofer catalogue, while demonstrating robust effectiveness, revealed both strengths and limitations in practical application. Its exhaustive and detailed nature ensures comprehensive coverage, but is time intensive. This thoroughness, while beneficial for certification rigour, needs to be balanced against practical time constraints in real-world scenarios.
The evaluation of alternative catalogues provided additional insights. The T\"UV catalogue's incomplete publication status rendered it unsuitable for standalone certification efforts. The Auditing Machine Learning Algorithms catalogue showed promise for certification purposes, potentially offering a more streamlined approach compared to the Fraunhofer methodology. However, its less structured nature suggests a need for deeper AI system expertise. But it might potentially be a faster certification process while maintaining quality standards.

\subsubsection{Limitations}
Several key limitations emerged during the certification process. The absence of an active development team is limiting, as it prevented the implementation of the typical feedback loop essential for certification refinements. This limitation transformed the certification process into evaluation rather than an iterative improvement process, highlighting the importance of ongoing development support for successful certification efforts.
Documentation gaps cannot be addressed through subsequent submissions. This emphasized the importance of comprehensive initial documentation of the chosen system. 

Any actors performing a future sample certification of systems that are not actively developed should keep this in mind, and they should choose a system with the most comprehensive documentation.

\subsubsection{Recommendations for future Sample Certifications}
The experience gained from this study suggests several crucial considerations for future certification efforts. For AI application selection, auditors should identify AI systems that exist within a broader application setting with surrounding code. The AI system itself, as well as the application setting and surrounding code, should be extensively documented. The ideal certification candidate should have an active development team willing to engage in the certification process.

A practical recommendation that emerged from this study was the value of creating a centralized archive of all available information and documentation before initiating the certification process.

\subsubsection{Recommendations for Real-World Applications}
Our findings yield several practical recommendations for real-world certification implementations. The Fraunhofer catalogue, while highly detailed and extensive, requires significant time investment for thorough completion. However, its precision and comprehensiveness make it a valuable tool for certification processes. Particularly noteworthy are the initial sections of the catalogue, specifically the AI lifecycle overview, which prove especially effective in providing auditors with comprehensive insights into the AI system's general functionality. This initial overview serves as an excellent starting point for any certification process. Although we did not extensively examine the Auditing Machine Learning Algorithms Catalogue in this work, its different structural approach suggests potential for more efficient certification processes. It could offer auditors a faster path to system certification while maintaining appropriate quality.

\section{Conclusion and Future Research Directions}
\label{sec7}

This work-in-progress presents our ongoing research into the practical application of AI certification catalogues, providing insights into both the capabilities and limitations of current certification approaches, as well as the challenges we encountered in certifying an open-source AI system. The implementation of the Fraunhofer AI Assessment Catalogue demonstrated its effectiveness as a comprehensive certification tool, particularly in its systematic approach to evaluate AI systems. We also found that the approach is, at times, bulky and time-consuming in some areas.

In future works, other approaches and certification catalogues should be considered, to potentially streamline the process. The certification process highlights the critical importance of complete system documentation and active engagement from the development team. The absence of these elements can negatively impact the certification process and even make certification infeasible. During the certification attempt, we identified the key characteristics an AI system must possess to be adequately certified. Before selecting a public AI system for a sample certification, auditors should consider factors like thorough documentation and accessibility to the development team. These findings directly address the initial research objectives by identifying both useful aspects of this certification approach and areas for improvement. They also provide practical insights into the certification process and its limitations. Future work in this field could focus on developing more flexible certification methodologies that accommodate various system states and development scenarios. 

\vspace{2mm} \noindent \textbf{Acknowledgements:} This work was supported by the FFG COMET program, and the strategic COMET project ``KnowCERTIFAI'' led by Know Center Research GmbH.


\bibliography{biblography_main}


\begin{thebibliography}{45}
\ifx \bisbn   \undefined \def \bisbn  #1{ISBN #1}\fi
\ifx \binits  \undefined \def \binits#1{#1}\fi
\ifx \bauthor  \undefined \def \bauthor#1{#1}\fi
\ifx \batitle  \undefined \def \batitle#1{#1}\fi
\ifx \bjtitle  \undefined \def \bjtitle#1{#1}\fi
\ifx \bvolume  \undefined \def \bvolume#1{\textbf{#1}}\fi
\ifx \byear  \undefined \def \byear#1{#1}\fi
\ifx \bissue  \undefined \def \bissue#1{#1}\fi
\ifx \bfpage  \undefined \def \bfpage#1{#1}\fi
\ifx \blpage  \undefined \def \blpage #1{#1}\fi
\ifx \burl  \undefined \def \burl#1{\textsf{#1}}\fi
\ifx \doiurl  \undefined \def \doiurl#1{\url{https://doi.org/#1}}\fi
\ifx \betal  \undefined \def \betal{\textit{et al.}}\fi
\ifx \binstitute  \undefined \def \binstitute#1{#1}\fi
\ifx \binstitutionaled  \undefined \def \binstitutionaled#1{#1}\fi
\ifx \bctitle  \undefined \def \bctitle#1{#1}\fi
\ifx \beditor  \undefined \def \beditor#1{#1}\fi
\ifx \bpublisher  \undefined \def \bpublisher#1{#1}\fi
\ifx \bbtitle  \undefined \def \bbtitle#1{#1}\fi
\ifx \bedition  \undefined \def \bedition#1{#1}\fi
\ifx \bseriesno  \undefined \def \bseriesno#1{#1}\fi
\ifx \blocation  \undefined \def \blocation#1{#1}\fi
\ifx \bsertitle  \undefined \def \bsertitle#1{#1}\fi
\ifx \bsnm \undefined \def \bsnm#1{#1}\fi
\ifx \bsuffix \undefined \def \bsuffix#1{#1}\fi
\ifx \bparticle \undefined \def \bparticle#1{#1}\fi
\ifx \barticle \undefined \def \barticle#1{#1}\fi
\bibcommenthead
\ifx \bconfdate \undefined \def \bconfdate #1{#1}\fi
\ifx \botherref \undefined \def \botherref #1{#1}\fi
\ifx \url \undefined \def \url#1{\textsf{#1}}\fi
\ifx \bchapter \undefined \def \bchapter#1{#1}\fi
\ifx \bbook \undefined \def \bbook#1{#1}\fi
\ifx \bcomment \undefined \def \bcomment#1{#1}\fi
\ifx \oauthor \undefined \def \oauthor#1{#1}\fi
\ifx \citeauthoryear \undefined \def \citeauthoryear#1{#1}\fi
\ifx \endbibitem  \undefined \def \endbibitem {}\fi
\ifx \bconflocation  \undefined \def \bconflocation#1{#1}\fi
\ifx \arxivurl  \undefined \def \arxivurl#1{\textsf{#1}}\fi
\csname PreBibitemsHook\endcsname

\bibitem[\protect\citeauthoryear{Costa and Aparicio}{2023}]{costa_applications_2023}
\begin{barticle}
\bauthor{\bsnm{Costa}, \binits{C.J.}},
\bauthor{\bsnm{Aparicio}, \binits{M.}}:
\batitle{Applications of {Data} {Science} and {Artificial} {Intelligence}}.
\bjtitle{Applied Sciences}
\bvolume{13}(\bissue{15}),
\bfpage{9015}
(\byear{2023})
\doiurl{10.3390/app13159015} .
\bcomment{Number: 15 Publisher: Multidisciplinary Digital Publishing Institute}
\end{barticle}
\endbibitem

\bibitem[\protect\citeauthoryear{Rajpurkar et~al.}{2022}]{rajpurkar2022ai}
\begin{barticle}
\bauthor{\bsnm{Rajpurkar}, \binits{P.}},
\bauthor{\bsnm{Chen}, \binits{E.}},
\bauthor{\bsnm{Banerjee}, \binits{O.}},
\bauthor{\bsnm{Topol}, \binits{E.J.}}:
\batitle{Ai in health and medicine}.
\bjtitle{Nature medicine}
\bvolume{28}(\bissue{1}),
\bfpage{31}--\blpage{38}
(\byear{2022})
\end{barticle}
\endbibitem

\bibitem[\protect\citeauthoryear{Van~den Broek et~al.}{2021}]{van2021machine}
\begin{botherref}
\oauthor{\bsnm{Broek}, \binits{E.}},
\oauthor{\bsnm{Sergeeva}, \binits{A.}},
\oauthor{\bsnm{Huysman}, \binits{M.}}:
When the machine meets the expert: An ethnography of developing ai for hiring.
MIS quarterly
\textbf{45}(3)
(2021)
\end{botherref}
\endbibitem

\bibitem[\protect\citeauthoryear{Cao}{2022}]{cao2022ai}
\begin{barticle}
\bauthor{\bsnm{Cao}, \binits{L.}}:
\batitle{Ai in finance: challenges, techniques, and opportunities}.
\bjtitle{ACM Computing Surveys (CSUR)}
\bvolume{55}(\bissue{3}),
\bfpage{1}--\blpage{38}
(\byear{2022})
\end{barticle}
\endbibitem

\bibitem[\protect\citeauthoryear{Kowald et~al.}{2013}]{kowald2013social}
\begin{bchapter}
\bauthor{\bsnm{Kowald}, \binits{D.}},
\bauthor{\bsnm{Dennerlein}, \binits{S.}},
\bauthor{\bsnm{Theiler}, \binits{D.}},
\bauthor{\bsnm{Walk}, \binits{S.}},
\bauthor{\bsnm{Trattner}, \binits{C.}}:
\bctitle{The social semantic server: A framework to provide services on social semantic network data}.
In: \bbtitle{9th International Conference on Semantic Systems, I-SEMANTICS 2013},
pp. \bfpage{50}--\blpage{54}
(\byear{2013}).
\bcomment{CEUR}
\end{bchapter}
\endbibitem

\bibitem[\protect\citeauthoryear{Lacic et~al.}{2014a}]{lacic2014recommending}
\begin{bchapter}
\bauthor{\bsnm{Lacic}, \binits{E.}},
\bauthor{\bsnm{Kowald}, \binits{D.}},
\bauthor{\bsnm{Seitlinger}, \binits{P.C.}},
\bauthor{\bsnm{Trattner}, \binits{C.}},
\bauthor{\bsnm{Parra}, \binits{D.}}:
\bctitle{Recommending items in social tagging systems using tag and time information}.
In: \bbtitle{In Proceedings of the 1st Social Personalization Workshop Co-located with Hypertext'14},
pp. \bfpage{4}--\blpage{9}
(\byear{2014}).
\bcomment{Association of Computing Machinery}
\end{bchapter}
\endbibitem

\bibitem[\protect\citeauthoryear{Lacic et~al.}{2014b}]{lacic2014towards}
\begin{bchapter}
\bauthor{\bsnm{Lacic}, \binits{E.}},
\bauthor{\bsnm{Kowald}, \binits{D.}},
\bauthor{\bsnm{Parra}, \binits{D.}},
\bauthor{\bsnm{Kahr}, \binits{M.}},
\bauthor{\bsnm{Trattner}, \binits{C.}}:
\bctitle{Towards a scalable social recommender engine for online marketplaces: The case of apache solr}.
In: \bbtitle{Proceedings of the 23rd International Conference on World Wide Web},
pp. \bfpage{817}--\blpage{822}
(\byear{2014})
\end{bchapter}
\endbibitem

\bibitem[\protect\citeauthoryear{Lacic et~al.}{2015}]{lacic2015utilizing}
\begin{bchapter}
\bauthor{\bsnm{Lacic}, \binits{E.}},
\bauthor{\bsnm{Kowald}, \binits{D.}},
\bauthor{\bsnm{Eberhard}, \binits{L.}},
\bauthor{\bsnm{Trattner}, \binits{C.}},
\bauthor{\bsnm{Parra}, \binits{D.}},
\bauthor{\bsnm{Marinho}, \binits{L.B.}}:
\bctitle{Utilizing online social network and location-based data to recommend products and categories in online marketplaces}.
In: \bbtitle{Mining, Modeling, and Recommending'Things' in Social Media: 4th International Workshops, MUSE 2013, Prague, Czech Republic, September 23, 2013, and MSM 2013, Paris, France, May 1, 2013, Revised Selected Papers},
pp. \bfpage{96}--\blpage{115}
(\byear{2015}).
\bcomment{Springer}
\end{bchapter}
\endbibitem

\bibitem[\protect\citeauthoryear{Kasinidou et~al.}{2024}]{kasinidou_artificial_2024}
\begin{bchapter}
\bauthor{\bsnm{Kasinidou}, \binits{M.}},
\bauthor{\bsnm{Kleanthous}, \binits{S.}},
\bauthor{\bsnm{Busso}, \binits{M.}},
\bauthor{\bsnm{Rodas}, \binits{M.}},
\bauthor{\bsnm{Otterbacher}, \binits{J.}},
\bauthor{\bsnm{Giunchiglia}, \binits{F.}}:
\bctitle{Artificial {Intelligence} in {Everyday} {Life} 2.0: {Educating} {University} {Students} from {Different} {Majors}}.
In: \bbtitle{Proceedings of the 2024 on {Innovation} and {Technology} in {Computer} {Science} {Education} {V}. 1}.
\bsertitle{{ITiCSE} 2024},
pp. \bfpage{24}--\blpage{30},
\bconflocation{New York, NY, USA}
(\byear{2024})
\end{bchapter}
\endbibitem

\bibitem[\protect\citeauthoryear{Baum et~al.}{2023}]{baum_fear_2023}
\begin{botherref}
\oauthor{\bsnm{Baum}, \binits{K.}},
\oauthor{\bsnm{Bryson}, \binits{J.}},
\oauthor{\bsnm{Dignum}, \binits{F.}},
\oauthor{\bsnm{Dignum}, \binits{V.}},
\oauthor{\bsnm{Grobelnik}, \binits{M.}},
\oauthor{\bsnm{Hoos}, \binits{H.}},
\oauthor{\bsnm{Irgens}, \binits{M.}},
\oauthor{\bsnm{Lukowicz}, \binits{P.}},
\oauthor{\bsnm{Muller}, \binits{C.}},
\oauthor{\bsnm{Rossi}, \binits{F.}},
\oauthor{\bsnm{Shawe-Taylor}, \binits{J.}},
\oauthor{\bsnm{Theodorou}, \binits{A.}},
\oauthor{\bsnm{Vinuesa}, \binits{R.}}:
From fear to action: {AI} governance and opportunities for all.
Frontiers in Computer Science
\textbf{5}
(2023).
Publisher: Frontiers
\end{botherref}
\endbibitem

\bibitem[\protect\citeauthoryear{Kowald et~al.}{2024}]{kowald2024establishing}
\begin{barticle}
\bauthor{\bsnm{Kowald}, \binits{D.}},
\bauthor{\bsnm{Scher}, \binits{S.}},
\bauthor{\bsnm{Pammer-Schindler}, \binits{V.}},
\bauthor{\bsnm{M{\"u}llner}, \binits{P.}},
\bauthor{\bsnm{Waxnegger}, \binits{K.}},
\bauthor{\bsnm{Demelius}, \binits{L.}},
\bauthor{\bsnm{Fessl}, \binits{A.}},
\bauthor{\bsnm{Toller}, \binits{M.}},
\bauthor{\bsnm{Mendoza~Estrada}, \binits{I.G.}},
\bauthor{\bsnm{{\v{S}}imi{\'c}}, \binits{I.}}, \betal:
\batitle{Establishing and evaluating trustworthy ai: overview and research challenges}.
\bjtitle{Frontiers in Big Data}
\bvolume{7},
\bfpage{1467222}
(\byear{2024})
\end{barticle}
\endbibitem

\bibitem[\protect\citeauthoryear{Smuha}{2021}]{smuha_race_2021}
\begin{barticle}
\bauthor{\bsnm{Smuha}, \binits{N.A.}}:
\batitle{From a ‘race to {AI}’ to a ‘race to {AI} regulation’: regulatory competition for artificial intelligence}.
\bjtitle{Law, Innovation and Technology}
\bvolume{13}(\bissue{1}),
\bfpage{57}--\blpage{84}
(\byear{2021})
\doiurl{10.1080/17579961.2021.1898300} .
\bcomment{Publisher: Routledge \_eprint: https://doi.org/10.1080/17579961.2021.1898300}.
Accessed 2024-10-28
\end{barticle}
\endbibitem

\bibitem[\protect\citeauthoryear{Mueck et~al.}{2022}]{mueck_etsi_2022}
\begin{botherref}
\oauthor{\bsnm{Mueck}, \binits{M.}},
\oauthor{\bsnm{Cadzow}, \binits{S.}},
\oauthor{\bsnm{Communications}, \binits{C.}},
\oauthor{\bsnm{Wood}, \binits{S.}}:
{ETSI} {Activities} in the field of {Artificial} {Intelligence} {Preparing} the implementation of the {European} {AI} {Act} - 1st {Edition} – {December} -2022.
Technical report,
ETSI
(2022)
\end{botherref}
\endbibitem

\bibitem[\protect\citeauthoryear{Nad et~al.}{2023}]{nad_trustworthiness_2023}
\begin{botherref}
\oauthor{\bsnm{Nad}, \binits{T.}},
\oauthor{\bsnm{Scher}, \binits{S.}},
\oauthor{\bsnm{Königstorfer}, \binits{F.}}:
Trustworthiness of {AI}.
Technical Report NIST AI 100-1,
SGS
(June 2023)
\end{botherref}
\endbibitem

\bibitem[\protect\citeauthoryear{Winter et~al.}{2021}]{winter_trusted_2021}
\begin{botherref}
\oauthor{\bsnm{Winter}, \binits{P.M.}},
\oauthor{\bsnm{Eder}, \binits{S.}},
\oauthor{\bsnm{Weissenböck}, \binits{J.}},
\oauthor{\bsnm{Schwald}, \binits{C.}},
\oauthor{\bsnm{Doms}, \binits{T.}},
\oauthor{\bsnm{Vogt}, \binits{T.}},
\oauthor{\bsnm{Hochreiter}, \binits{S.}},
\oauthor{\bsnm{Nessler}, \binits{B.}}:
Trusted {Artificial} {Intelligence}: {Towards} {Certification} of {Machine} {Learning} {Applications}.
arXiv.
arXiv:2103.16910 [cs, stat]
(2021)
\end{botherref}
\endbibitem

\bibitem[\protect\citeauthoryear{Poretschkin et~al.}{2023}]{poretschkin_ai_2023}
\begin{botherref}
\oauthor{\bsnm{Poretschkin}, \binits{D.M.}},
\oauthor{\bsnm{Schmitz}, \binits{A.}},
\oauthor{\bsnm{Akila}, \binits{D.M.}},
\oauthor{\bsnm{Adilova}, \binits{L.}},
\oauthor{\bsnm{Becker}, \binits{D.D.}},
\oauthor{\bsnm{Cremers}, \binits{D.A.B.}},
\oauthor{\bsnm{Hecker}, \binits{D.D.}},
\oauthor{\bsnm{Houben}, \binits{D.S.}},
\oauthor{\bsnm{Rosenzweig}, \binits{J.}},
\oauthor{\bsnm{Sicking}, \binits{J.}},
\oauthor{\bsnm{Schulz}, \binits{E.}},
\oauthor{\bsnm{Voss}, \binits{D.A.}},
\oauthor{\bsnm{Wrobel}, \binits{D.S.}}:
{AI} {Assessment} {Catalog}.
Technical report,
Fraunhofer IAIS
(February 2023)
\end{botherref}
\endbibitem

\bibitem[\protect\citeauthoryear{Pimentel}{2024}]{pimentel_why_2024}
\begin{botherref}
\oauthor{\bsnm{Pimentel}, \binits{B.}}:
Why {AI} still needs regulation despite impact
(2024).
\url{https://legal.thomsonreuters.com/blog/why-ai-still-needs-regulation-despite-impact/}
Accessed 2024-10-28
\end{botherref}
\endbibitem

\bibitem[\protect\citeauthoryear{Benjamens et~al.}{2020}]{benjamens_state_2020}
\begin{botherref}
\oauthor{\bsnm{Benjamens}, \binits{S.}},
\oauthor{\bsnm{Dhunnoo}, \binits{P.}},
\oauthor{\bsnm{Mesko}, \binits{B.}}:
The state of artificial intelligence-based {FDA}-approved medical devices and algorithms: an online database.
npj Digital Medicine
\textbf{3}
(2020)
\doiurl{10.1038/s41746-020-00324-0}
\end{botherref}
\endbibitem

\bibitem[\protect\citeauthoryear{Saenz et~al.}{2023}]{saenz_autonomous_2023}
\begin{barticle}
\bauthor{\bsnm{Saenz}, \binits{A.D.}},
\bauthor{\bsnm{Harned}, \binits{Z.}},
\bauthor{\bsnm{Banerjee}, \binits{O.}},
\bauthor{\bsnm{Abràmoff}, \binits{M.D.}},
\bauthor{\bsnm{Rajpurkar}, \binits{P.}}:
\batitle{Autonomous {AI} systems in the face of liability, regulations and costs}.
\bjtitle{npj Digital Medicine}
\bvolume{6}(\bissue{1}),
\bfpage{1}--\blpage{3}
(\byear{2023})
\doiurl{10.1038/s41746-023-00929-1} .
\bcomment{Publisher: Nature Publishing Group}.
Accessed 2024-06-01
\end{barticle}
\endbibitem

\bibitem[\protect\citeauthoryear{European-Union}{2024}]{european-union_regulation_2024}
\begin{botherref}
\oauthor{\bsnm{European-Union}}:
Regulation ({EU}) 2024/1689 of the {European} {Parliament} and of the {Council} of 13 {June} 2024 laying down harmonised rules on artificial intelligence and amending {Regulations} ({EC}) {No} 300/2008, ({EU}) {No} 167/2013, ({EU}) {No} 168/2013, ({EU}) 2018/858, ({EU}) 2018/1139 and ({EU}) 2019/2144 and {Directives} 2014/90/{EU}, ({EU}) 2016/797 and ({EU}) 2020/1828 ({Artificial} {Intelligence} {Act}){Text} with {EEA} relevance.
Legislative Body: CONSIL, EP
(2024).
\url{http://data.europa.eu/eli/reg/2024/1689/oj/eng}
Accessed 2024-07-20
\end{botherref}
\endbibitem

\bibitem[\protect\citeauthoryear{Fernhout and Duquin}{2024}]{fernhout_eu_2024}
\begin{botherref}
\oauthor{\bsnm{Fernhout}, \binits{F.}},
\oauthor{\bsnm{Duquin}, \binits{T.}}:
The {EU} {Artificial} {Intelligence} {Act}: our 16 key takeaways {\textbar} {Stibbe}
(2024).
\url{https://www.stibbe.com/publications-and-insights/the-eu-artificial-intelligence-act-our-16-key-takeaways}
Accessed 2024-08-07
\end{botherref}
\endbibitem

\bibitem[\protect\citeauthoryear{Madiega}{2023}]{madiega_artificial_2023}
\begin{botherref}
\oauthor{\bsnm{Madiega}, \binits{T.}}:
Artificial intelligence liability directive
(2023)
\end{botherref}
\endbibitem

\bibitem[\protect\citeauthoryear{{European-Union}}{2022}]{european-union_proposal_2022}
\begin{botherref}
\oauthor{\bsnm{{European-Union}}}:
Proposal for a {DIRECTIVE} {OF} {THE} {EUROPEAN} {PARLIAMENT} {AND} {OF} {THE} {COUNCIL} on adapting non-contractual civil liability rules to artificial intelligence ({AI} {Liability} {Directive})
(2022).
\url{https://eur-lex.europa.eu/legal-content/EN/TXT/?uri=CELEX%3A52022PC0496}
Accessed 2024-11-16
\end{botherref}
\endbibitem

\bibitem[\protect\citeauthoryear{White-House}{2022}]{white-house_blueprint_2022}
\begin{botherref}
\oauthor{\bsnm{White-House}}:
Blueprint for an {AI} {Bill} of {Rights}
(2022).
\url{https://www.whitehouse.gov/wp-content/uploads/2022/10/Blueprint-for-an-AI-Bill-of-Rights.pdf}
\end{botherref}
\endbibitem

\bibitem[\protect\citeauthoryear{Department~for Digital}{2022}]{department_for_digital_culture_media__sport_establishing_2022}
\begin{botherref}
\oauthor{\bsnm{Digital}, \binits{M..S.} \bsuffix{Culture}}:
Establishing a pro-innovation approach to regulating {AI}
(2022).
\url{https://www.gov.uk/government/publications/establishing-a-pro-innovation-approach-to-regulating-ai/establishing-a-pro-innovation-approach-to-regulating-ai-policy-statement}
Accessed 2024-06-02
\end{botherref}
\endbibitem

\bibitem[\protect\citeauthoryear{Blösser and Weihrauch}{2023}]{blosser_consumer_2023}
\begin{barticle}
\bauthor{\bsnm{Blösser}, \binits{M.}},
\bauthor{\bsnm{Weihrauch}, \binits{A.}}:
\batitle{A consumer perspective of {AI} certification – the current certification landscape, consumer approval and directions for future research}.
\bjtitle{European Journal of Marketing}
\bvolume{58}(\bissue{2}),
\bfpage{441}--\blpage{470}
(\byear{2023}).
\bcomment{Publisher: Emerald Publishing Limited}
\end{barticle}
\endbibitem

\bibitem[\protect\citeauthoryear{Delgado-Aguilera~Jurado et~al.}{2024}]{delgado-aguilera_jurado_introduction_2024}
\begin{barticle}
\bauthor{\bsnm{Delgado-Aguilera~Jurado}, \binits{R.}},
\bauthor{\bsnm{Ye}, \binits{X.}},
\bauthor{\bsnm{Ortolá~Plaza}, \binits{V.}},
\bauthor{\bsnm{Zamarreño~Suárez}, \binits{M.}},
\bauthor{\bsnm{Pérez~Moreno}, \binits{F.}},
\bauthor{\bsnm{Arnaldo~Valdés}, \binits{R.M.}}:
\batitle{An introduction to the current state of standardization and certification on military {AI} applications}.
\bjtitle{Journal of Air Transport Management}
\bvolume{121},
\bfpage{102685}
(\byear{2024})
\end{barticle}
\endbibitem

\bibitem[\protect\citeauthoryear{{Benjamin Fresz} et~al.}{2024}]{benjamin_fresz_contribution_2024}
\begin{botherref}
\oauthor{\bsnm{{Benjamin Fresz}}},
\oauthor{\bsnm{{Vincent Philipp Göbels}}},
\oauthor{\bsnm{{Safa Omri}}},
\oauthor{\bsnm{{Danilo Brajovic}}}:
The {Contribution} of {XAI} for the {Safe} {Development} and {Certification} of {AI}: {An} {Expert}-{Based} {Analysis}
(2024).
\url{https://arxiv.org/html/2408.02379v1}
Accessed 2024-11-16
\end{botherref}
\endbibitem

\bibitem[\protect\citeauthoryear{{Hadrien Pouget}}{2024}]{hadrien_pouget_standardsetzung_2024}
\begin{botherref}
\oauthor{\bsnm{{Hadrien Pouget}}}:
Standardsetzung {\textbar} {EU}-{Gesetz} zur künstlichen {Intelligenz}
(2024).
\url{https://artificialintelligenceact.eu/de/standardeinstellung/}
Accessed 2024-11-16
\end{botherref}
\endbibitem

\bibitem[\protect\citeauthoryear{SAI-FI-DE-NL-NO-UK}{2023}]{sai-fi-de-nl-no-uk_auditing_2023}
\begin{botherref}
\oauthor{\bsnm{SAI-FI-DE-NL-NO-UK}}:
Auditing machine learning algorithms.
Technical report,
Supreme Audit Institutions FI, DE, NL, NO, UK
(February 2023)
\end{botherref}
\endbibitem

\bibitem[\protect\citeauthoryear{{Angelica Perez} et~al.}{2021}]{angelica_perez_thoughtworksartsemopy_2021}
\begin{botherref}
\oauthor{\bsnm{{Angelica Perez}}},
\oauthor{\bsnm{{Julien Deswaef}}},
\oauthor{\bsnm{{Puneetha Pai}}}:
thoughtworksarts/{EmoPy}.
Thoughtworks Arts.
original-date: 2017-12-20T02:19:22Z
(2021).
\url{https://github.com/thoughtworksarts/EmoPy}
Accessed 2024-07-23
\end{botherref}
\endbibitem

\bibitem[\protect\citeauthoryear{Thoughtworks}{2018}]{thoughtworks_riot_2018}
\begin{botherref}
\oauthor{\bsnm{Thoughtworks}}:
{RIOT} {\textbar} {Thoughtworks} {Arts}
(2018).
\url{https://thoughtworksarts.io/projects/riot/}
Accessed 2024-10-29
\end{botherref}
\endbibitem

\bibitem[\protect\citeauthoryear{Semmelrock et~al.}{2023}]{semmelrock2023reproducibility}
\begin{botherref}
\oauthor{\bsnm{Semmelrock}, \binits{H.}},
\oauthor{\bsnm{Kopeinik}, \binits{S.}},
\oauthor{\bsnm{Theiler}, \binits{D.}},
\oauthor{\bsnm{Ross-Hellauer}, \binits{T.}},
\oauthor{\bsnm{Kowald}, \binits{D.}}:
Reproducibility in machine learning-driven research.
arXiv preprint arXiv:2307.10320
(2023)
\end{botherref}
\endbibitem

\bibitem[\protect\citeauthoryear{Semmelrock et~al.}{2024}]{semmelrock2024reproducibility}
\begin{botherref}
\oauthor{\bsnm{Semmelrock}, \binits{H.}},
\oauthor{\bsnm{Ross-Hellauer}, \binits{T.}},
\oauthor{\bsnm{Kopeinik}, \binits{S.}},
\oauthor{\bsnm{Theiler}, \binits{D.}},
\oauthor{\bsnm{Haberl}, \binits{A.}},
\oauthor{\bsnm{Thalmann}, \binits{S.}},
\oauthor{\bsnm{Kowald}, \binits{D.}}:
Reproducibility in machine learning-based research: Overview, barriers and drivers.
arXiv preprint arXiv:2406.14325
(2024)
\end{botherref}
\endbibitem

\bibitem[\protect\citeauthoryear{Perez}{2018}]{perez_emopy_2018}
\begin{botherref}
\oauthor{\bsnm{Perez}, \binits{A.}}:
{EmoPy}: a machine learning toolkit for emotional expression
(2018).
\url{https://www.thoughtworks.com/insights/blog/emopy-machine-learning-toolkit-emotional-expression}
Accessed 2024-10-29
\end{botherref}
\endbibitem

\bibitem[\protect\citeauthoryear{Palmer}{2016}]{palmer_riot_2016}
\begin{botherref}
\oauthor{\bsnm{Palmer}, \binits{K.}}:
{RIOT} {AI}
(2016).
\url{http://karenpalmer.uk/portfolio/riot/}
Accessed 2024-10-29
\end{botherref}
\endbibitem

\bibitem[\protect\citeauthoryear{Thoughtworks}{2019}]{thoughtworks_thoughtworksartsriot_2019}
\begin{botherref}
\oauthor{\bsnm{Thoughtworks}}:
thoughtworksarts/riot.
Thoughtworks Arts.
original-date: 2017-11-22T15:59:17Z
(2019).
\url{https://github.com/thoughtworksarts/riot}
Accessed 2024-10-29
\end{botherref}
\endbibitem

\bibitem[\protect\citeauthoryear{{TED Residency}}{2018}]{ted_residency_karen_2018}
\begin{botherref}
\oauthor{\bsnm{{TED Residency}}}:
Karen {Palmer}: {The} film that watches you back
(2018).
\url{https://www.youtube.com/watch?v=Rw8gLEkFdSw}
Accessed 2024-10-29
\end{botherref}
\endbibitem

\bibitem[\protect\citeauthoryear{Thoughtworks}{2018}]{thoughtworks_thoughtworks_2018}
\begin{botherref}
\oauthor{\bsnm{Thoughtworks}}:
Thoughtworks {Arts} {Exhibition} at {SPRING}/{BREAK} for {Armory} {Week} 2018 {\textbar} {Thoughtworks} {Arts}
(2018).
\url{https://thoughtworksarts.io/blog/thoughtworks-arts-exhibition-spring-break-armory-week/}
Accessed 2024-10-29
\end{botherref}
\endbibitem

\bibitem[\protect\citeauthoryear{{Thoughtworks}}{2017}]{thoughtworks_karen_2017}
\begin{botherref}
\oauthor{\bsnm{{Thoughtworks}}}:
Karen {Palmer} {Awarded} {Thoughtworks} {AI} {Residency} {\textbar} {Thoughtworks} {Arts}
(2017).
\url{https://thoughtworksarts.io/blog/karen-palmer-ai-residency/}
Accessed 2024-10-29
\end{botherref}
\endbibitem

\bibitem[\protect\citeauthoryear{{Karen Palmer}}{2017}]{karen_palmer_riot_2017}
\begin{botherref}
\oauthor{\bsnm{{Karen Palmer}}}:
{RIOT} {Video}
(2017).
\url{https://www.youtube.com/watch?v=-BCny9SuI3A}
Accessed 2024-10-29
\end{botherref}
\endbibitem

\bibitem[\protect\citeauthoryear{Perez}{2018}]{perez_recognizing_2018}
\begin{botherref}
\oauthor{\bsnm{Perez}, \binits{A.}}:
Recognizing human facial expressions with machine learning
(2018).
\url{https://www.thoughtworks.com/insights/articles/recognizing-human-facial-expressions-machine-learning}
Accessed 2024-10-29
\end{botherref}
\endbibitem

\bibitem[\protect\citeauthoryear{Team}{2017}]{emopy_development_team_welcome_2017}
\begin{botherref}
\oauthor{\bsnm{Team}, \binits{E.D.}}:
Welcome to {EmoPy}’s documentation! — {EmoPy} 1.0 documentation
(2017).
\url{https://emopy.readthedocs.io/en/latest/}
Accessed 2024-10-29
\end{botherref}
\endbibitem

\bibitem[\protect\citeauthoryear{Microsoft}{2023}]{microsoft_microsoftferplus_2023}
\begin{botherref}
\oauthor{\bsnm{Microsoft}}:
microsoft/{FERPlus}.
Microsoft.
original-date: 2016-09-14T06:35:21Z
(2023).
\url{https://github.com/microsoft/FERPlus}
Accessed 2024-10-29
\end{botherref}
\endbibitem

\bibitem[\protect\citeauthoryear{Cohn}{2024}]{cohn_resources_2024}
\begin{botherref}
\oauthor{\bsnm{Cohn}, \binits{J.}}:
Resources – {Jeffrey} {Cohn}
(2024).
\url{https://www.jeffcohn.net/resources/}
Accessed 2024-10-29
\end{botherref}
\endbibitem

\end{thebibliography}

\end{document}